# Intelligent GPS Spoofing Attack Detection in Power Grids

M. Sabouri, S. Siamak, M. Dehghani, M. Mohammadi, M.H Asemani

Applied Control & Robotic Research Laboratory, Shiraz University, Shiraz, Iran, m.sabouri@shirazu.ac.ir

School of Electrical and Computer Engineering, Shiraz, Iran, s.siyamak@shirazu.ac.ir

School of Electrical and Computer Engineering, Shiraz, Iran, mdehghani@shirazu.ac.ir

School of Mechanical Engineering Shiraz University, Shiraz, Iran, mohsen_mohammadi@shirazu.ac.ir

Applied Control & Robotic Research Laboratory, Shiraz University, Shiraz, Iran, asemani@shirazu.ac.ir

*Abstract*— **The GPS is vulnerable to GPS spoofing attack (GSA), which leads to disorder in time and position results of the GPS receiver. In power grids, phasor measurement units (PMUs) use GPS to build time-tagged measurements, so they are susceptible to this attack. As a result of this attack, sampling time and phase angle of the PMU measurements change. In this paper, a neural network GPS spoofing detection (NNGSD) with employing PMU data from the dynamic power system is presented to detect GSAs. Numerical results in different conditions show the real-time performance of the proposed detection method**

## I. Introduction

Since GPS systems use wireless communication, receivers are vulnerable to cyberattacks, including GPS spoofing attacks. The spoofer generates false GPS signals and broadcasts them [1]. The adjacent GPS receiver tracks transmitted fake signals, so false time tags are received [2].

With the development of power grids into microgrids and their complexity, the use of PMUs as voltage and current phasor metering sensors is expanding [3]–[5]. The unique feature of them is network synchronization because they use accurate GPS time for sampling simultaneously [6].

Recent research in the field of detecting PMU attacks in power systems reveals two distinct approaches. These two categories can be divided into the following topics. 1-Model-dependent approaches 2- Learning-based approaches.

Given the importance of GPS spoofing attacks that lead to power system errors, it has always been a matter of significance to detect attacks as fast as possible which lead to a lot of research in this area. Among the model-dependent investigations, the following can be mentioned: In [7] an approach to investigate the vulnerability of PMUs in counterfeiting attacks and attack reconstruction is offered. In another study, two methods are presented to compare the information received from GPS receiver antennas with a specific pattern for attack detection [8]. In [9], a static model is developed to detect synchrophasor-based system states while a GSA is happened. In [10], an algorithm of integrating phasor measurement methods and state estimation methods to detect attacks is presented. Moreover, in [11], a correction system for states based on comparison with the measured states are presented for false data under attack.

In model-based approaches, the main focus is on applying a system model to estimate the states or phase angles of the system, while in learning-based methods, data is of paramount importance [12]–[16]. In [17], using deep learning methods, FDI attacks are studied, and the behavioral characteristics of the power system against FDI attacks are extracted from past system data that utilize these features to detect attacks. [18] uses machine learning algorithms to classify secure and under FDI attacks measurements. In [19], a method is presented to detect anomalies based on FDI attacks using an observer consisting of two parts, Luenberger, and a neural network.

In this paper, an artificial neural network(ANN)-based approach is proposed to detect GPS spoofing attacks using information measured by the PMU. Unlike methods that require updating their detectors, the proposed method can only perform one-time basic training of attack detection operations with less complexity and computation time. The performance of this method is such that it can be used even in various conditions such as network load changes and in the presence of noise. Moreover, the provided detector has the capability of detecting GPS spoofing attacks under different conditions and on several PMUs, despite the simplicity of the structure. The idea is based on training an ANN by a variety of attack models using different affected datasets of attacks.

The structure of this article is organized as follows. Section II introduces the basic concepts related to the structure of a power grid as well as basic information on ANNs. Section III discusses the approach used to detect GPS attacks. In Section IV, the simulation results of the proposed method for several attacks are discussed. Finally, the results and conclusions of this work are discussed in the final section.

## II. Preliminary Issues

### A. Power System Model

A general linear dynamic model of a large scale power system is as follows [3], [4]:

$$\dot{x}_i(t) = A_i x_i(t) + B_i u_i(t) + \sum_{\substack{j=1 \\ j \neq i}}^{N} H_{ij} x_j$$
$$+ w_i(t) \qquad , \qquad i = 1 \dots N_g \tag{1}$$

$$z_i(t) = C_i x_i + \sum_{\substack{j=1 \\ j \neq i}}^{N} L_{ij} x_j + \xi_i(t), \tag{2}$$

where $N_g$ is the number of subsystems. $A_i$ and $H_{ij}$ represent the transfer matrices of each subsystem individually and the relationship of each subsystem to the other subsystems, respectively. $C_i$ and $L_{ij}$ are the matrices of the measurement of each subsystem and the relationship of each subsystem to other subsystems and $w_i$ and $\xi_i$ are the noise vectors.

### B. Neural Network Model

ANNs are one of the areas of machine learning which have different types based on the architecture and the type of training algorithm. Feedforward backpropagation Neural Network (FNN) is a class of neural networks commonly used to model or classify and cluster training data. The knowledge extracted from the data by these neural networks will be used later for prediction purposes.

The FNN is designed to find a hidden pattern or a nonlinear relationship between the input arguments and the output arguments (target) [20]. An ANN is made up of a large number of special interconnected processing elements called neurons; each captures the output of the previous layer and applies a specific weight and activation function, then transfers to the next layer [21] as follows:

$$Y(x) = f\left( b + \sum_{i=1}^{n} x_i w_j \right) \tag{3}$$

where $Y$ is the neural network function with $R^z \to R^l$ mapping. $z$ and $l$ are the number of input and output, respectively. $x_i$ is the $i$th row of neural network input, $w_j$ is the $j$th column of weight matrix, $b$ is the network bias vector and $n$ is the number of samples [22].

### C. GPS Spoofing Attack Model in Power Grid

The effect of GSA on the PMU measurements is a phase shift that is proportional to the amount of time manipulated by the spoofer. This variation is same for all measurement signals at the time and it does not affect the absolute of the signals [2]. The relation between the spoofed measurement data and correct data is defined by (4) and (5):

$$x(t) = |x(t)| \sin(\omega t + \theta) \tag{4}$$

$$x^{spf}(t) = |x(t)| \sin(\omega t + \theta + \theta_{spf}) \tag{5}$$

where $x(t)$ is a signal with phase angle $\theta$ without any attacks. $x^{spf}(t)$ is the spoofed $x(t)$ and $\theta_{spf}$ is the change in the phase of the measured signal after GPS spoofing.

### III. PROPOSED DETECTION ALGORITHM

This section describes the method used to detect GSA. The process of NNGSD is illustrated in Fig. 1. It is based on a multilayer ANN trained in a specific process once and then, by receiving the real-time PMU measurements, NNGSD can detect the occurrence and location of GSAs. The structure of NNGSD consists of the following sections.

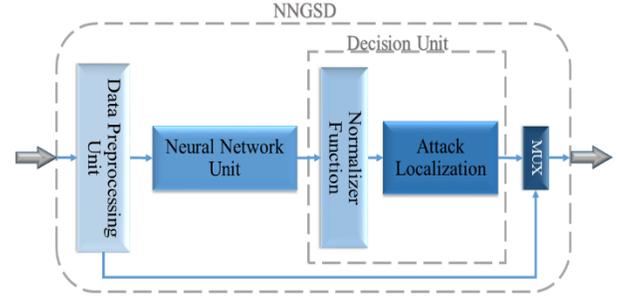

Figure 1. Neural network GPS spoofing detector (NNGSD) structure

### A. Preprocessing Unit

This unit handles the initial processing required to prepare the input data of the neural network. The input data of this block is considered PMU measurements which consist of rotor angles of generator buses in the power system. This block works in two stages. At first, it separates the time tags from the data measured by the PMUs and then, performs a preliminary analysis of the data obtained to detect and filter the clean data.

### B. Neural Network Unit

The neural network used in the NNGSD structure is a multilayer perceptron (MLP) neural network which can be created by expanding the standard form of neural network in (3). Considering $N$ as the hidden layers number plus input and output layers. Fig. 2 shows the structure of a multilayer network similar to the network used in this study.

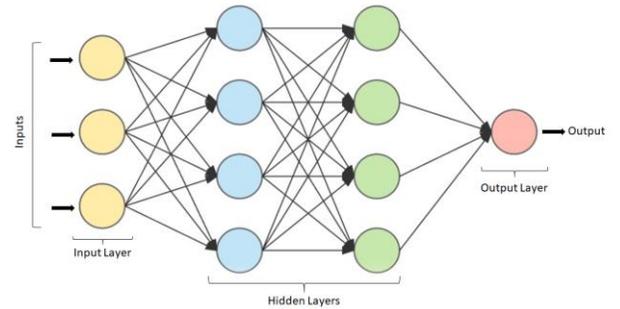

Figure 2. Neural network

The process of gathering data to train the neural network is as follows: Attacks are applied on the power grid with a specific framework and the data measured by the PMUs with sufficient information on how and where the attack occurs is formulated into the matrix $M$ in (6):

$$M = \begin{bmatrix} x_{ij} & y_{ij} \end{bmatrix} \tag{6}$$

where $n$ is the number of samples. Each element of $x_{ij}$ is the angle of $j$th rotor at $i$th sample. $y_{ij}$ is a member of set $\{0,1\}$ where 1 indicates the occurrence of the attack and 0 represents the non-occurrence of an attack. By determining the matrix $M$ and the inputs and the outputs of the training data using (1) and (2), the neural network is trained and the matrices $W$ and $b$ of each layer are calculated.

### C. Decision Unit

This unit provides the last decision about the detection of GSA. The decision unit consists of the normalizer function and attack localization blocks. The objective of normalizer function block is to normalize the output of NN and to facilitate the attack locating process. The following equation is used to calculate the output vector of the normalizer block at the $i$th sample time and $j$th output:

$$S(y_{ij}) = u(y_{ij} - \alpha) = \begin{cases} 1 & y_{ij} \geq \alpha \\ 0 & y_{ij} < \alpha \end{cases} \quad (7)$$

where $u(.)$ is a step function and $\alpha$ is the normalization threshold coefficient which in addition to increasing the accuracy of the neural network output, improves the output of data analysis in subsequent blocks. The output of the normalizer unit $S(y_i)$ is given to the attack localization unit as an input vector. In this unit, if some elements of the input vector are one, the associated PMUs are reported as being attacked. The zero elements reveal the security of the related PMUs are. Fig. 3 shows the normalizer function block.

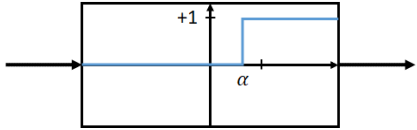

Figure 3. Normalizer function block

Finally, the Mux block tags the time data to the output data of attack location block.

## IV. SIMULATION RESULTS

In this section, the proposed method in section III is implemented and simulated in MATLAB software for the standard IEEE-14 bus power system with 5 generators. The simulation and verification process of the spoof detector is as follows.

The neural network construction applied in the spoofing detector is defined based on the MLP neural network. The number of hidden layers is considered 3 and therefore, N=5. The set of $S_q = \{z, 20, 50, 20, l\}$ shows the sequence of neurons number in each layer. In this set, $z = 5$ and $l = 5$. The $f_1$, $f_2$, $f_3$ and $f_4$ are considered sigmoid functions with formula $y = e^x/(e^x + 1)$. Fig. 4 shows the implemented neural network in MATLAB.

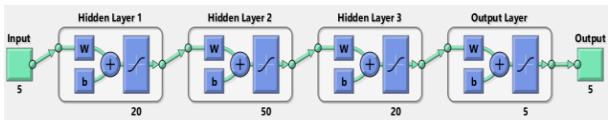

Figure 4. Neural network diagram

For neural network training, in the normal condition of the power system, numerous spoofing attacks are applied to the system deliberately and by simulating (1) and (2), resulting in the input and output data for $n = 10^4$ cases. Then, according to the data, matrix $M$ is constituted. The output of the neural network for each PMU under spoofing attack is 1 and otherwise, it is 0. In this paper, the number of PMUs is equal to the number of generators of power system ($N_g = 5$). The length of computation time window is 10 seconds. The parameters of the neural network are initially considered as standard and are calculated by trial and error parameters. By forming training matrices, neural network inputs and outputs are integrated into MATLAB training process.

In the learning process of neural network using MATLAB software, the training data are automatically divided into three categories of training, testing and validation and the software presents a report on the accuracy of data training in these three sections. The overall accuracy of the proposed neural network at the training process is 98.4117%.

To test the proposed neural network (NN) and NNGSD, new GSAs are considered as shown in Fig. 5 and they are tested in three conditions; normal operation, in the presence of noise and in load change condition. The outputs are calculated by applying the input data to the NNGSD. Then, the accuracy of the proposed detector is tested by comparing the output of the detector to the simulated output data.

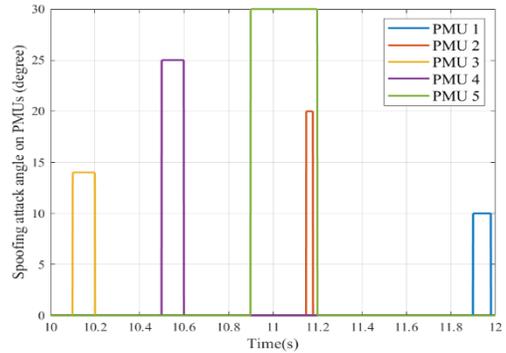

Fig. 5. The angle of GSAs considered on PMUs

### A. GSA Detection in the Normal Operation of Power System

In this case, the performance of the NNGSD under normal and standard condition of the power system considering some GSAs on different PMUs are investigated. The results of the NNGSD outputs are shown in Fig. 6.

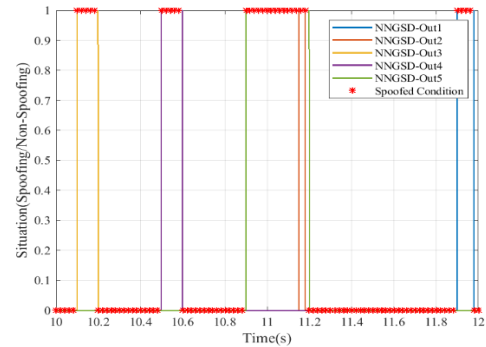

Fig. 6. Detection of multi GSAs on the PMUs

As shown in Fig. 6, the results of proposed method are with high precision. Table 2 shows the detection percentage of proposed NN and NNGSD.

Table 2. Percentage of attack detection on PMUs in normal operation

| | Normal Measurement + GSA attack | |
|---|---|---|
| | NN Output | NNGSD |
| PMU1 | 91.9691 | 95.7534 |
| PMU2 | 92.1414 | 97.3672 |
| PMU3 | 92.9975 | 98.8955 |
| PMU4 | 92.9990 | 98.9701 |
| PMU5 | 92.9973 | 98.0152 |
| **Overall** | **92.6208** | **98.8002** |

### B. GSA Detection in the Presence of Noise

To test the robustness of the NNGSD and its performance in the presence of noise, some further analysis is done. In noisy condition, noise signals with normal distribution function and variance $\sigma = 0.1$ with SNR $\cong$ 20 db are added to the measured data in simulation. By applying the noisy data, which also includes the effects of the spoofing attack, the spoofing detection results of NN and NNGSD are shown in Fig. 7 and 8, respectively. The attack detection percentage of NN and NNGSD are given in Table 3.

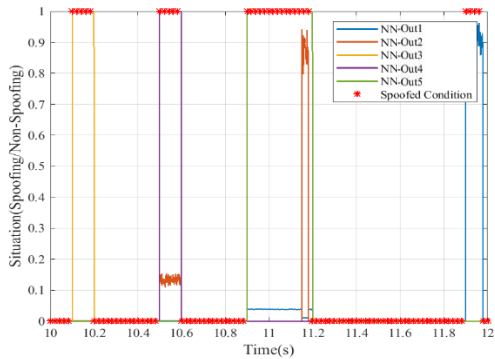

Fig. 7. Detection of multi GSA on PMUs in the presence of noise using NN

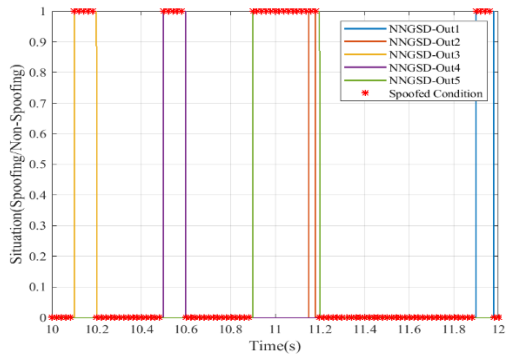

Fig. 8. Detection of multi GSA on the PMUs in the presence of noise using NNGSD

Table 3. Percentage of attack detection on PMUs in presence of noise

| | Noisy Measurement + GSA attack | |
|---|---|---|
| | NN Output | NNGSD |
| PMU1 | 89.0123 | 92.1088 |
| PMU2 | 91.5009 | 97.6895 |
| PMU3 | 91.9975 | 97.9234 |
| PMU4 | 91.9990 | 98.1308 |
| PMU5 | 91.9973 | 98.2525 |
| **Overall** | **91.3014** | **96.8210** |

### C. GSA Detection in the Load Changing Condition

In the following, the existence of controlled power load changes is considered. In this case, the load changes are applied to the measured data in the presence of the attack by changing the input of the control in (1). The outputs of NN and NNGSD are shown in Fig. 9 and 10, respectively. The GSA detection percentage of NN and NNGSD with measurements in the presence of noise and load changing are shown in Table 4.

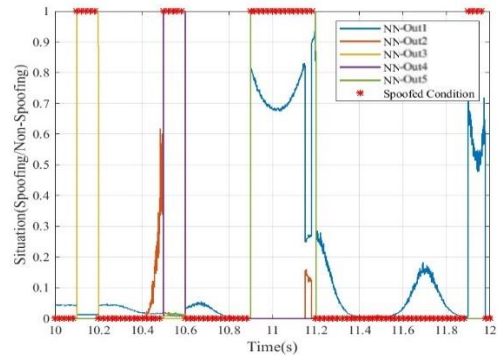

Fig. 9. The output of NN with spoofed measurements in the presence of noise and load changing

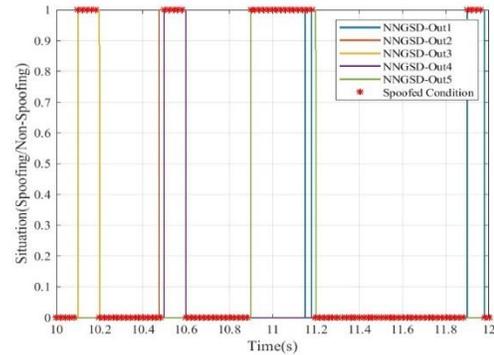

Fig. 10. The output of NNGSD with spoofed measurements in the presence of noise and load changing

Table 4. Percentage of attack detection on PMUs

| | Measurements under load changing + GSA attack | | Measurements under noise and load changing + GSA attack | |
|---|---|---|---|---|
| | NN Output | NNGSD | NN Output | NNGSD |
| PMU1 | 77.5918 | 86.5067 | 77.5811 | 85.2569 |
| PMU2 | 90.2368 | 97.2514 | 87.7733 | 91.2509 |
| PMU3 | 90.9976 | 98.9725 | 91.9976 | 99. 8925 |
| PMU4 | 90.9975 | 98.9881 | 91.9975 | 99.9912 |
| PMU5 | 90.9301 | 98.9976 | 91.9284 | 99.8714 |
| **Overall** | **88.1507** | **96.1432** | **88.2555** | **94.0926** |

## V. Conclusion

In this paper, a GPS spoofing detector based on a neural network is proposed. The structure of the NNGSD consists of two main parts; a trained neural network and a decision block component which announce the final diagnosis about the GPS spoofing attacks and the location of them. The proposed method is tested by noisy and in load changing conditions which haven't been previously in the training process.

Simulation results indicate the efficacy of the proposed method in real-time GSA detecting while the measurements are noisy or the system loads are changed suddenly.